\documentclass[twocolumn,amsmath,aps]{revtex4}
\pdfoutput=1

\usepackage{amssymb}
\usepackage{amsmath}
\usepackage{epsfig}
\usepackage{subfigure}
\usepackage{mathrsfs}
\usepackage{longtable}
\usepackage{hyperref}
\usepackage[retainorgcmds]{IEEEtrantools}

\begin{document}

\title{Cosmology with Eddington-inspired Gravity}
\author{James H. C. Scargill}
\email{james.scargill@physics.ox.ac.uk}
\affiliation{Theoretical Physics, University of Oxford, Rudolf Peierls Centre, 1 Keble Road, Oxford, OX1 3NP, UK}
\author{M\'aximo Banados}
 \email{maxbanados@fis.puc.cl}
 \affiliation{P. Universidad Cat\'olica de Chile, Avenida Vicuna Mackema 4860,
 Santiago, Chile}
\author{Pedro G. Ferreira}
\email{p.ferreira1@physics.ox.ac.uk}
\affiliation{Astrophysics, University of Oxford, Denys Wilkinson Building, Keble Road, Oxford, OX1 3RH, UK}
\begin{abstract}
    We study the dynamics of homogeneous, isotropic universes which are governed by the Eddington-inspired alternative theory of gravity which has a single extra parameter, $\kappa$. Previous results showing singularity-avoiding behaviour for $\kappa > 0$ are found to be upheld in the case of domination by a perfect fluid with equation of state parameter $w > 0$. The range $-\frac{1}{3} < w < 0$ is found to lead to universes which experience unbounded expansion rate whilst still at a finite density. In the case $\kappa < 0$ the addition of spatial curvature is shown to lead to the possibility of oscillation between two finite densities. Domination by a scalar field with an exponential potential is found to also lead to singularity-avoiding behaviour when $\kappa > 0$. Certain values of the parameters governing the potential lead to behaviour in which the expansion rate of the universe changes sign several times before transitioning to regular GR-like behaviour.
\end{abstract}

\maketitle

\section{Introduction}

In 1924  Eddington proposed a theory of gravity based solely on a connection field, without introducing a metric. A particular extension of Eddignton's theory including  matter fields was recently considered in \cite{Banados and Ferreira} (see also \cite{Vollick}). In a vacuum, this theory reproduces `Eddington-inspired' (or just `Eddington') gravity and is completely equivalent to General Relativity (GR), a fact which should be contrasted with other modified gravity theories in which vacuum deviations are expected (e.g. metric $f(R)$ theories \cite{vacuum deviations}).  On the other hand deviations from GR are seen in regions of high density.  Where one would expect singularities such as the interior of black holes or the birth of the universe, in Eddington inspired gravity often these are avoided \cite{Eddington stars, Banados and Ferreira, Avelino and Ferreira} (although other singularities may appear in an astrophysical context  \cite{Pani+Sotiriou}).
The appearance of such behaviour in a purely classical theory such as Eddington gravity is interesting and exciting.

The cosmological implications of this theory were initially considered in \cite{Banados and Ferreira}, and studied in more depth in \cite{Avelino and Ferreira} and \cite{Cho et al}; it was found that universes containing ordinary matter avoid singularites, with the exact behaviour depending on the sign of one of the parameters of the theory, and being either a regular bounce (from contraction to expansion) or `loitering' around a minimum size before transitioning to GR-like expansion. It has also recently been shown that
tensor perturbations may render the regular behaviour of such cosmological models unstable near the minimum scale (either in the asymptotic past or at the bounce \cite{Celia}).

In this paper we confirm those findings (in the addition of spatial curvature to the Friedmann-Robertson-Walker (FRW) metric), and present other interesting behaviour viz. universes which experience unbounded expansion rate at finite density akin to undergoing a second big bang/`cosmic hiccup', and closed universes which oscillate between two finite values of the density. Finally we investigate the dynamics of universes in which the dominant contribution to the stress-energy tensor is due to a scalar field with an exponential potential, as has been done in the case of GR \cite{Ferreira+Joyce}.

This paper is structured as follows: section \ref{definitions} covers some basic definitions and the governing field equations; section \ref{Friedman eqn. general} exhibits the Friedman equation and qualitatively compares it to GR; section \ref{w/o phi} examines solutions to this with domination by one kind of perfect fluid; section \ref{k > 0, w > 0} demonstrates the `loitering' behaviour which has previously been observed; section \ref{k > 0, -1/3 < w < 0} investigates the `cosmic hiccup' mentioned above; section \ref{oscillating universes} exhibits the oscillating universes; section \ref{w phi} covers the case of domination by a scalar field; section \ref{loitering} investigates the `loitering' behaviour which has been observed for the case of regular matter fields and which shows up again in the case of scalar fields; section \ref{expconexp} exhibits some universes which show qualitatively new behaviour - expansion, followed by contraction, followed by expansion. In section \ref{conclusion} we conclude.

\subsection{Definitions and Field Equations} \label{definitions}

The action governing Eddington gravity is \cite{Banados and Ferreira}
\begin{align}
S[g,\Gamma,\Psi] = &\frac{2}{\kappa} \int \mathrm{d}^4 x \left[ \sqrt{| g + \kappa R(\Gamma) |} - \lambda \sqrt{| g |} \right]
+ S_M[g,\Psi], \label{Edd}
\end{align}
(throughout this paper we use reduced Planck units, $c = 8\pi G = 1$) where $R(\Gamma)$ denotes the symmetric part of the Ricci tensor, and $\lambda$ is a dimensionless constant which must satisfy $\lambda = 1 + \kappa \Lambda$ if GR is to be recovered in the limit of $\kappa R \ll g$ (though the cosmological constant will be assumed negligible in the following). Gravitational theories with such a Born-Infeld-like structure have previously been investigated in \cite{Deser}. Matter can, of course, be coupled to the connection as well as the metric, however for simplicity here it is assumed to only couple to $g$. Two field equations are derived from this action:
\begin{align}
q_{\mu\nu} &= g_{\mu\nu} + \kappa R_{\mu\nu}(q), \label{1st fe} \\
\sqrt{| q |} q^{\mu\nu} - \lambda \sqrt{| g |} g^{\mu\nu} &= - \kappa \sqrt{| g |} T^{\mu\nu}, \label{2nd fe}
\end{align}
in which $q$ is an \emph{auxiliary metric} which is compatible with the connection i.e. $\Gamma^{\alpha}_{\beta\gamma} = \frac{1}{2} q^{\alpha\nu}(\partial_\beta q_{\nu\gamma} + \partial_\gamma q_{\beta\nu} - \partial_\nu q_{\beta\gamma})$, and $q^{\mu\nu} \equiv [q^{-1}]^{\mu\nu}$.

The action (\ref{Edd}), and its equations of motion (\ref{1st fe},\ref{2nd fe}), can also be derived from the following bigravity theory,
\begin{eqnarray}\label{actiongq}
I[g,q] &=& \int \sqrt{q} (q^{\mu\nu}R_{\mu\nu}(q) - 2\Lambda ) \\  & & \ \ + \   \Lambda (\sqrt{q}q^{\mu\nu} g_{\mu\nu} - 2\sqrt{g} ) + \sqrt{g} L_{m}(g)\nonumber
\end{eqnarray}
with
\begin{equation}\label{}
\Lambda = {1 \over \kappa}
\end{equation}
Varying (\ref{actiongq}) with respect to $q_{\mu\nu}$ and $g_{\mu\nu}$ one obtains a set of equations fully equivalent to (\ref{1st fe},\ref{2nd fe}). This form of the action is particularly interesting because it shows a remarkable close relation with the recently discovered family of unitary massive gravity theories \cite{dRGT,Hassan+Rosen}. Indeed, massive gravities are built as bigravity theories where the potential is a linear combination of the symmetric polynomials of the eigenvalues of the matrix \begin{equation}\label{}
\gamma = \sqrt{q^{-1} g }.
\end{equation}
For this class of potentials the theory has no Boulware-Deser instability and, in particular, the massive sector carries 5 degrees of freedom instead of 6.  Besides the cosmological constants (which also appear in (\ref{actiongq})), one such symmetric polynomial is
\begin{equation}\label{}
\sqrt{q}\left( \mbox{Tr}(\gamma^2) - (\mbox{Tr}\gamma)^2 \right) = \sqrt{q} ( q^{\mu\nu}g_{\mu\nu} -  (\mbox{Tr}\,\gamma)^2 )
\end{equation}
The first term is precisely the interaction between $q_{\mu\nu}$ and $g_{\mu\nu}$ appearing in (\ref{actiongq}), resulting in Eddington theory coupled to matter. The second piece (more difficult to compute because the square root has to be taken before the trace) is not part of the Eddington theory. We find it intriguing that Eddington's theory coupled to matter is so close to the recently discovered stable massive gravity theories. We shall exploit this connection elsewhere.

\subsection{Friedman Equation} \label{Friedman eqn. general}

The FRW metric is \begin{equation}
ds^2 = -\mathrm{d}t^2 + a(t)^2 \left(\frac{\mathrm{d}r^2}{1-kr^2} + r^2( \mathrm{d}\theta^2 + \mathrm{sin}^2 \theta\, \mathrm{d}\phi^2)\right), \label{FRW}
\end{equation}
and we take the auxiliary metric to be
\begin{align}
q_{\mu\nu} \mathrm{d}x^{\mu} \mathrm{d}x^{\nu} = &-U(t)\mathrm{d}t^2 + a(t)^2 V(t) \nonumber \\
&\times \left(\frac{\mathrm{d}r^2}{1-k_q r^2} + r^2( \mathrm{d}\theta^2 + \mathrm{sin}^2 \theta\, \mathrm{d}\phi^2)\right) \label{auxiliary metric}
\end{align}
The energy momentum tensor is
\begin{equation}
T_{00} = \sum_i \rho_i + \rho_\phi \quad  \text{and} \quad  T_{ij} = a^2 \left(\sum_i P_i + P_\phi \right) \delta_{ij},
\end{equation}
where $\{i\}$ indicate the regular matter, and $\phi$ the scalar field. Finally
\begin{equation}
\rho_\phi = \frac{1}{2}\dot{\phi}^2 + V(\phi) \quad \text{and} \quad P_\phi = \frac{1}{2}\dot{\phi}^2 - V(\phi).
\end{equation}

The Eddington gravity field equations enforce $k_q = k$ and
 \begin{equation}
 U = \sqrt{\frac{(1-\kappa P_T)^3}{1+\kappa \rho_T}} \quad \text{and} \quad V = \sqrt{(1+\kappa \rho_T)(1-\kappa P_T)},
 \end{equation}
where $\rho_T = \sum_i \rho_i + \rho_\phi + \Lambda$ and $P_T = \sum_i P_i + P_\phi - \Lambda$.

The energy-momentum conservation equations are as in GR, so
\begin{align}
\ddot{\phi} + 3 H \dot{\phi} + V'(\phi) &= 0 \\
\dot{\rho}_i + 3 H \rho_i (1 + w_i) &= 0.
\end{align}

In the general case
\begin{equation}
H \equiv \frac{\dot{a}}{a} = \frac{1}{F} \left( \pm \sqrt{\frac{G}{6}} - B \right) \label{Friedman}
\end{equation}
with
\begin{align}
G &= \frac{1}{\kappa} \left[ 1 + 2U - 3 \frac{U}{V} - \frac{6 \kappa k}{a^2} \right] \\
F &= 1 - \bigg\{3\kappa \bigg[ \sum_i \left(1-w_i-\kappa(w_i \rho_T + P_T) \right) \rho_i (1 + w_i) \nonumber \\
 &\quad- \kappa(\rho_T + P_T) \dot{\phi}^2 \bigg] \bigg\} \bigg/ 4 (1+\kappa \rho_T)(1-\kappa P_T) \\
B &= \frac{1}{2}\kappa \frac{V' \dot{\phi}}{1 - \kappa P_T}. \label{B}
\end{align}

One key thing to note is that the scalar field does not simply enter in the same way as other forms of matter (as is the case in GR), which leads to the interesting result that although there are two solutions for $H$, they are not simply opposite signs, as in GR and Eddington gravity \emph{without} a scalar field.

If one assumes that $\mathrm{sign}(\dot{V}) = -\mathrm{sign}(H)$, i.e. as the universe increases in size, the potential energy of the scalar field decreases, then in fact the values for $H$ are simply of different signs. However whilst this assumption may hold for the low-density GR-like stage of evolution, it fails in the high density case with some interesting consequences as laid out in section \ref{expconexp}

Another departure from GR is that spatial curvature cannot be considered a fluid, since it does not enter in the same way as a fluid with $w = -\frac{1}{3}$ (which it would have to have in order to evolve as curvature).

Finally, in the low density and curvature limit ($\rho_i,\ \rho_\phi,\ \Lambda,\ k/a^2 \ll \kappa^{-1}$) GR is recovered to leading order:
\begin{widetext}
\begin{align}
H = &\pm\sqrt{\frac{\rho_T}{3} - \frac{k}{a^2}} \nonumber \\
&+ \kappa \bigg\{\pm \sqrt{\frac{\rho_T}{3} - \frac{k}{a^2}}\bigg[\frac{3}{4}\sum_i \rho_i (1-w_i^2) + \frac{1}{12}\left(\frac{\rho_T}{3} - \frac{k}{a^2}\right)^{-1} \left( \frac{15}{8} P_T^2 - \frac{9}{4} \rho_T P_T - \frac{17}{8} \rho_T^2 + \frac{6 k}{a^2}(P_T + \rho_T) \right) \bigg] -\frac{1}{2} V' \dot{\phi} \bigg\} \nonumber \\ &+ \mathcal{O}((\kappa \rho_T)^2).
\end{align}
\end{widetext}

\section{Dynamics of a universe without a scalar field} \label{w/o phi}

We first consider the dynamics of universes which do not posses scalar fields, have a negligible cosmological constant and, for simplicity, only one type of perfect fluid. The Friedman equation in this case is

\begin{align}
H^2 &= \frac{G}{6 F^2} \\ &= \frac{8}{3 \kappa} \Big[ (1+3w) \bar{\rho} - 2 + 2 \sqrt{(1+\bar{\rho})(1-w\bar{\rho})^3} \nonumber \\
&\quad\qquad - \frac{6 \bar{k}}{a^2}(1 - w \bar{\rho}) \Big] \nonumber \\
& \times \frac{(1+ \bar{\rho})(1 - w \bar{\rho})^2}{(4+(1-w)(1-3w)\bar{\rho}+2w(1+3w)\bar{\rho}^2)^2}, \label{Friedman w/o phi}
\end{align}
where $\bar{\rho} = \kappa \rho$ and $\bar{k} = \kappa k$.

Note that setting $w =-1$ (so that the universe contains only a cosmological constant) reduces eqn. (\ref{Friedman w/o phi}) to exactly that which would be expected in GR, as it should be since Eddington gravity exactly reproduces GR in a vacuum (with a cosmological constant).

\subsection{Behaviour of the Friedman equation}

The square root in the numerator restricts the domain of $H^2$ and the result is shown in table \ref{domain 1}.

\begin{table}[hbt]
\centering
\caption{Domain restrictions placed on $H^2$ by the square root}
\label{domain 1}
\begin{tabular}{crcl}
\hline \hline
$w$ & Valid &$\bar{\rho}$& range\\
\hline
$w > 0$ & $-1 \leq$ &$\bar{\rho}$& $\leq \frac{1}{w}$ \\
$w = 0$ & $-1 \leq$ &$\bar{\rho}$& \\
$-1 \leq w < 0$ & $ \bar{\rho} \leq \frac{1}{w} $&or&$ -1 \leq \bar{\rho}$ \\
$w \leq -1$ & $\bar{\rho} \leq -1 $&or&$ \frac{1}{w} \leq \bar{\rho}$ \\
\hline
\end{tabular}
\end{table}

The requirement $H^2 \geq 0$ fixes the sign of $\rho$ and table \ref{rho sign} shows the results (for $k$ = 0), from which we see that that the energy density is positive in all but two regions of parameter space. This is an interesting departure from GR which, in the spatially flat case, always requires non-negative energy densities.

\begin{table}[tb]
\centering
\caption{Sign of $\rho$ in various regions of parameter space, with no spatial curvature. $\bar{\rho}_{G*}$ satisfies $G(\bar{\rho}_{G*}) = H^2 (\bar{\rho}_{G*}) = 0$. ($w=-1$ is special but also has $\rho \geq 0$.)}
\label{rho sign}
\begin{tabular}{cccc}
\hline \hline
$w$ & $\bar{\rho}$ & sign($\kappa$) & sign($\rho$) \\
\hline
any & $0<\bar{\rho}$ & $+$ & $+$ \\
$-1 < w$ & $-1 < \bar{\rho} < 0$ & $-$ & $+$ \\
$-1<w<0$ & $\bar{\rho}_{G*} < \bar{\rho} < \frac{1}{w}$ & $+$ & $-$ \\
 & $\bar{\rho} < \bar{\rho}_{G*}$ & $-$ & $+$ \\
 $w < -1$ & $\bar{\rho}_{G*} < \bar{\rho} < 0$ & $-$ & $+$ \\
  & $\frac{1}{w} < \bar{\rho} < \bar{\rho}_{G*}$ & $+$ & $-$ \\
   & $\bar{\rho} < -1$ & $-$ & $+$ \\
\hline
\end{tabular}
\end{table}

As pointed out in \cite{Cho et al}, a final interesting quality of (\ref{Friedman w/o phi}) to note is that for certain ranges of $w$, $H^2 \to \infty$ for finite $\rho$. Section \ref{k > 0, -1/3 < w < 0} investigates a universe which exhibits such behaviour.

Looking at the discriminant of the quadratic in the denominator of (\ref{Friedman w/o phi}), we see that the discontinuities do not exist for $0.02 \approx \frac{1}{3}(7-4\sqrt{3}) < w < \frac{1}{3}(7+4\sqrt{3}) \approx 4.64$. Outside of this region we can determine graphically, given the constraints of table \ref{domain 1}, in which regions the discontinuities will appear (see appendix \ref{appendix discontinuities}), and the results are in table \ref{discontinuities}.

\begin{table*}[htb]
\centering
\caption{Regions in which discontinuities of $H^2$ appear and are not outside of the allowed range of $\bar{\rho}$.}
\label{discontinuities}
\begin{tabular}{ccccccc}
\hline \hline
$w < -1$ & $w = -1$ & $-1 < w < -\frac{1}{3}$ & $w = -\frac{1}{3}$ & $-\frac{1}{3} < w < 0$ & $0 \leq w < 4.64$... & $4.64... \leq w$ \\
\hline
$\bar{\rho}_{G*} < \bar{\rho} < 0$ & - & $\bar{\rho} < \bar{\rho}_{G*}$ & - & $0 < \bar{\rho}$ & - & $-1 < \bar{\rho} < 0$ \\
\hline
\end{tabular}
\end{table*}


Points at which $H^2 = 0$ and $\frac{\mathrm{d} H^2}{\mathrm{d} \rho} \neq 0$ will induce an expanding universe to collapse (and vice versa) since the universe cannot evolve further on this trajectory (because $H^2 \geq 0$) and the stationary point is unstable since only $H^2$, and not the sign of $H$, is fixed by the physics. Such `bounce' points exist in GR (if $k > 0$), and in Eddington gravity as well, i.e. the points $\bar{\rho} = -1$ and $\bar{\rho} = \bar{\rho}_{G*}$. We are not including $\rho = 0$ here because, although $H^2$ behaves as required, starting from a non-zero density this point cannot be reached in a finite time.

From (\ref{Friedman w/o phi}) we see that at $\bar{\rho} = w^{-1},\ H^2 = 0 \text{ and } \frac{\mathrm{d} H^2}{\mathrm{d} \rho} = 0$. This novel type of stationary point leads to the `loitering' behaviour remarked upon in \cite{Banados and Ferreira, Avelino and Ferreira}, and \cite{Cho et al}, in which the universe asymptotically approaches (or recedes from) a maximum density/minimum size. From table \ref{rho sign} we see that for a positive energy density, this type of universe will only exist for $w > 0 \text{ and } \kappa > 0$.

\subsection{The case $\kappa > 0$, $w > 0$} \label{k > 0, w > 0}

\begin{figure*}[htbp]
   \includegraphics[width=\columnwidth]{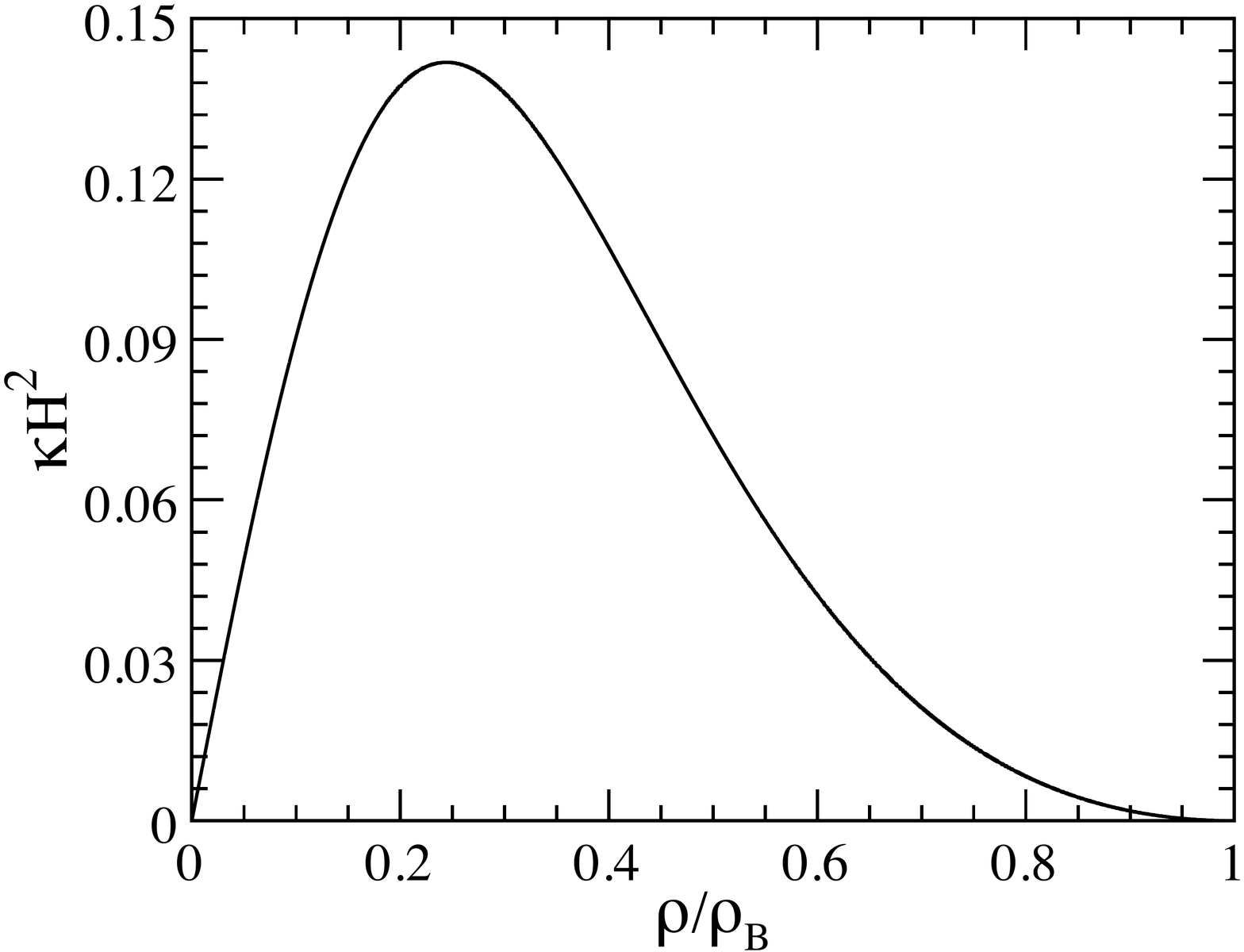}
   \includegraphics[width=\columnwidth]{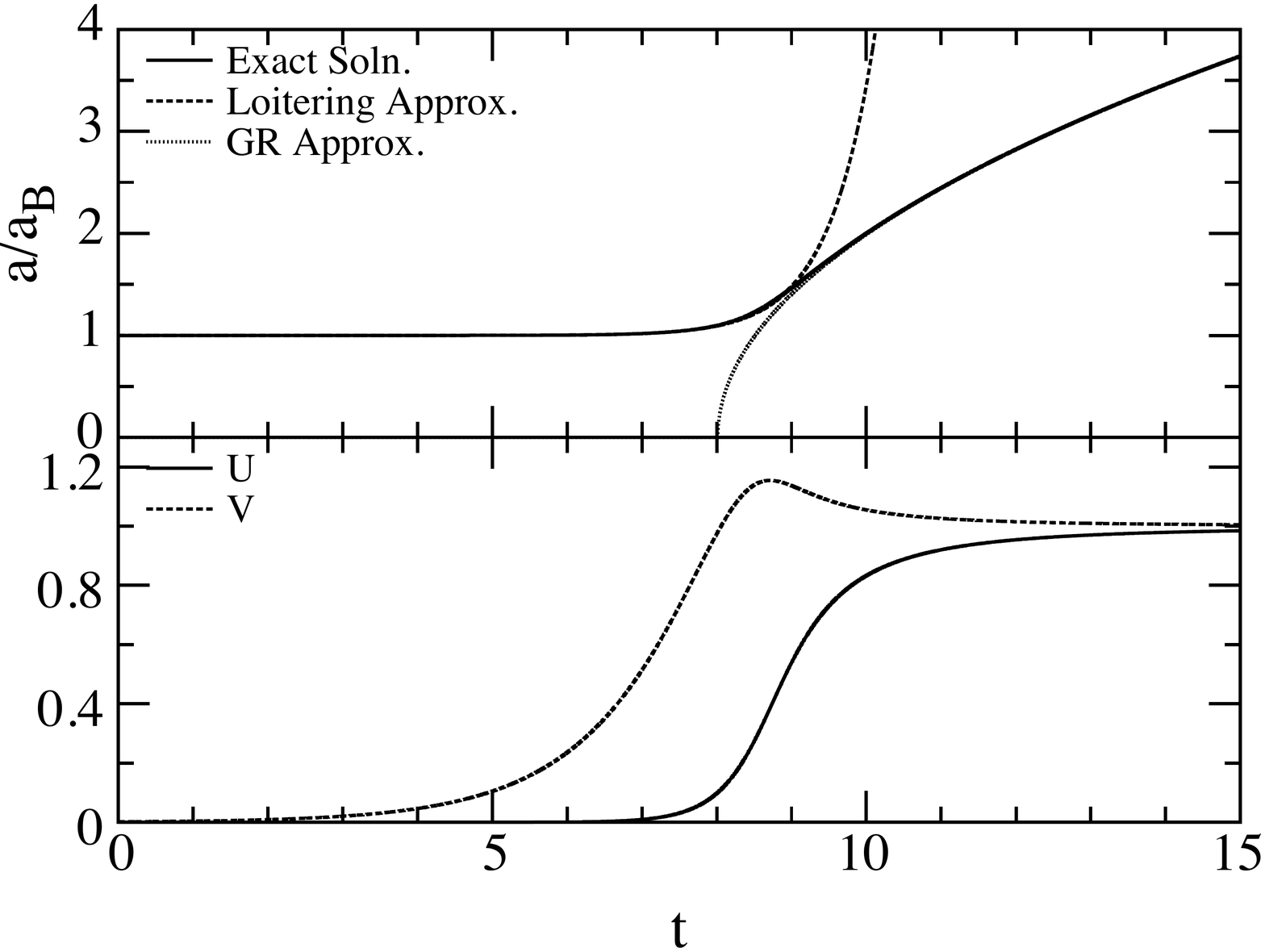}
      \caption{Left: Expansion rate of the universe against density (normalised by the maximum density $\rho_B = (\kappa w)^{-1}$), for $\kappa = 1$, $w = 1/3$, $k = 0$. \\
Right Top: Scale factor (normalised by the minimum scale factor $a_B$) against time (in reduced Planck units), for the full numerical solution to the Friedman equation, and approximations in two phases: loitering, as in section \ref{loitering phase}, and GR-like low-density limit. \\
   Right Bottom: Behaviour of the parameters describing the auxiliary metric (for the same universe).}
   \label{Hrho-aUV_1}
\end{figure*}

As seen in figure \ref{Hrho-aUV_1}, this type of universe exhibits the `loitering' behaviour mentioned in the previous section. We also see that to a good approximation the evolution can be split into two phases: the loitering phase in which deviations from GR are important, and the GR-like phase in which the dynamics of the expansion are approximately as given by GR.

\subsubsection{Loitering phase} \label{loitering phase}

Close to the maximum density $\bar{\rho} \equiv \bar{\rho}_B = w^{-1}$ the term involving $k$ is much smaller than the other terms in the numerator of (\ref{Friedman w/o phi}), and so we can neglect (small) spatial curvature during the loitering phase. Expanding about the maximum density, or equivalently the minimum scale factor $a_B$, to highest order the Friedman equation and deceleration parameter become
\begin{align}
H^2 &= \frac{8}{3\kappa} \left( \frac{w \delta \bar{\rho}}{3(1+w)} \right)^2 = \frac{8}{3\kappa}\left( \frac{\delta a}{a_B} \right)^2 \label{friedman loitering}\\
q_{dec} &= \left( \frac{w \delta \bar{\rho}}{3(1+w)} \right)^{-1} = - \left( \frac{\delta a}{a_B} \right)^{-1}
\end{align}
(where $\delta \bar{\rho} = \bar{\rho} - \bar{\rho}_B$ and $\delta a = a - a_B$). Using $H = \frac{\dot{a}}{a} = \frac{\dot{\delta a}}{a_B + \delta a}$, (\ref{friedman loitering}) can be solved to get
\begin{align}
\frac{a(t)}{a_B} &= 1 + \mathrm{exp}\left( \sqrt{\frac{8}{3\kappa}} (t-t_0) \right)\\
q_{dec} &= - \mathrm{exp}\left( - \sqrt{\frac{8}{3\kappa}} (t-t_0) \right).
\end{align}

Thus we see that the universe undergoes (accelerating) exponential expansion away from the minimum scale factor during this phase. The form of $a(t)$ is reminiscent of that observed in GR under domination by a cosmological constant, specifically with $\Lambda = 8 \kappa^{-1}$.

Calculating the particle horizon for such a universe we find
\begin{equation}
r_p(t) = \lim_{t_- \rightarrow - \infty} \int_{t_-}^t \frac{\mathrm{d} t'}{a(t')} = \lim_{t_- \rightarrow - \infty} \frac{t - t_-}{a_B}, \label{particle horizon Edd}
\end{equation}
which diverges. Again this is similar to GR under domination by a cosmological constant.

In \cite{Eddington stars} the authors note that Eddington gravity can be characterised as making the gravitational force repulsive at short distances (or high densities), which would agree with the behaviour which we see here. That is, the maximum density is reached as the repulsive effects of Eddington gravity oppose the traditional attractive effects of the matter density; which repulsive effects take a form similar to a cosmological constant with $\Lambda = 8 \kappa^{-1}$.

\subsubsection{Behaviour of the auxiliary metric}

From eqn. (\ref{auxiliary metric}) we see that the auxiliary metric $q_{\mu\nu}$ is identical to the metric $g_{\mu\nu}$ except that the temporal and spatial components of $q$ are modulated by the functions $U$ and $V$ respectively. Alternatively, by factoring out $U$, it can be considered to be related to an FRW metric with modified scale factor $a_q = a \sqrt{\frac{V}{U}}$, by an overall conformal transformation $U$. Thus $U$ and $V$, or $U$ and $a_q$ explain how the space described by the auxiliary metric behaves.

Figure \ref{Hrho-aUV_1} shows the behaviour of $U$, and $V$. As expected, in the GR regime $U,\ V \to 1$. On the other hand, close to the maximum density $U,\ V \to 0$, and the space described by $q$ becomes vanishingly small. Interestingly, $V$ reaches a maximum value just before the GR regime begins.

Expanding close to the maximum density we find
\begin{align}
U &= \sqrt{\frac{(3(1+w))^3}{2}} \exp\left( \frac{3}{2} \sqrt{\frac{8}{3\kappa}} (t-t_0) \right) \\
V &= \sqrt{\frac{3(1+w)}{2}} \exp\left( \frac{1}{2} \sqrt{\frac{8}{3\kappa}} (t-t_0) \right) \\
a_q &= \frac{a_B}{3(1+w)} \exp\left( - \frac{1}{2} \sqrt{\frac{8}{3\kappa}} (t-t_0) \right).
\end{align}

And so we see that, moving towards the maximum density, the temporal size decreases more rapidly than the spatial, with the result that $a_q$ actually goes infinite.

\subsubsection{The case $w = 0$} \label{k > 0, w = 0}

This case exhibits the same loitering behaviour at early times, but must be treated separately, since the density is unbounded; expanding about $a = 0$, to leading order we find:
\begin{align}
H^2 &= \frac{8}{3\kappa} \\
q_{dec} &= -1,
\end{align}
which is identical to GR with a cosmological constant $\Lambda = 8 \kappa^{-1}$. Therefore the particle horizon will diverge as in the case $w > 0$.

The behaviour of $U$ and $a_q$ is the same as in the $w > 0$ case, however in contrast, $V$ diverges as $t \to -\infty$. See appendix \ref{k > 0, w = 0, auxiliary metric} for the explicit functional forms.

\subsection{The case $\kappa > 0$, $-\frac{1}{3} < w < 0$} \label{k > 0, -1/3 < w < 0}

Firstly it should be noted that new physics (beyond Eddington gravity) is required in order to have a fluid with $w < 0$ dominating in the early universe; this problem is not considered here.

\begin{figure*}[htbp]
   \includegraphics[width=\columnwidth]{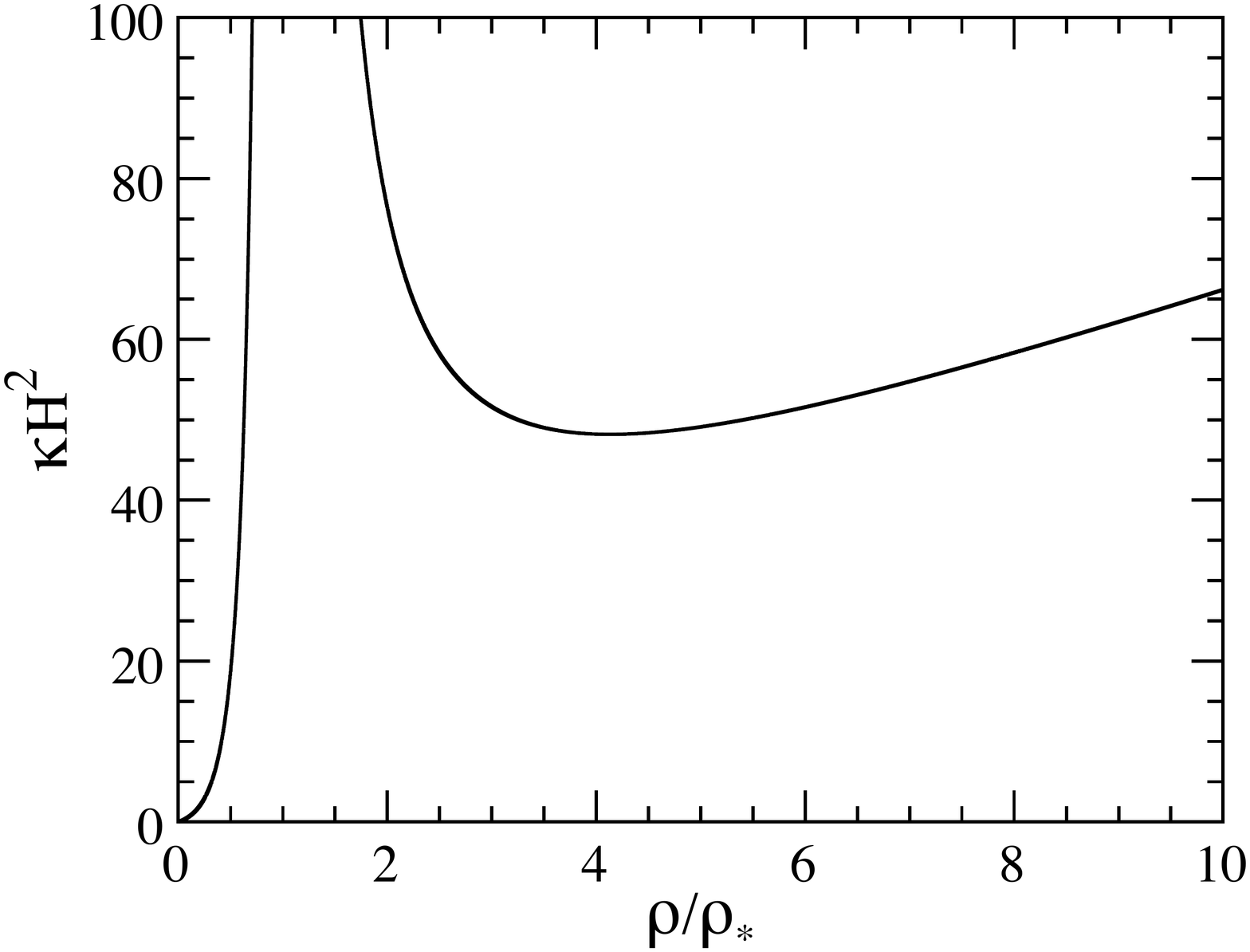}
   \includegraphics[width=\columnwidth]{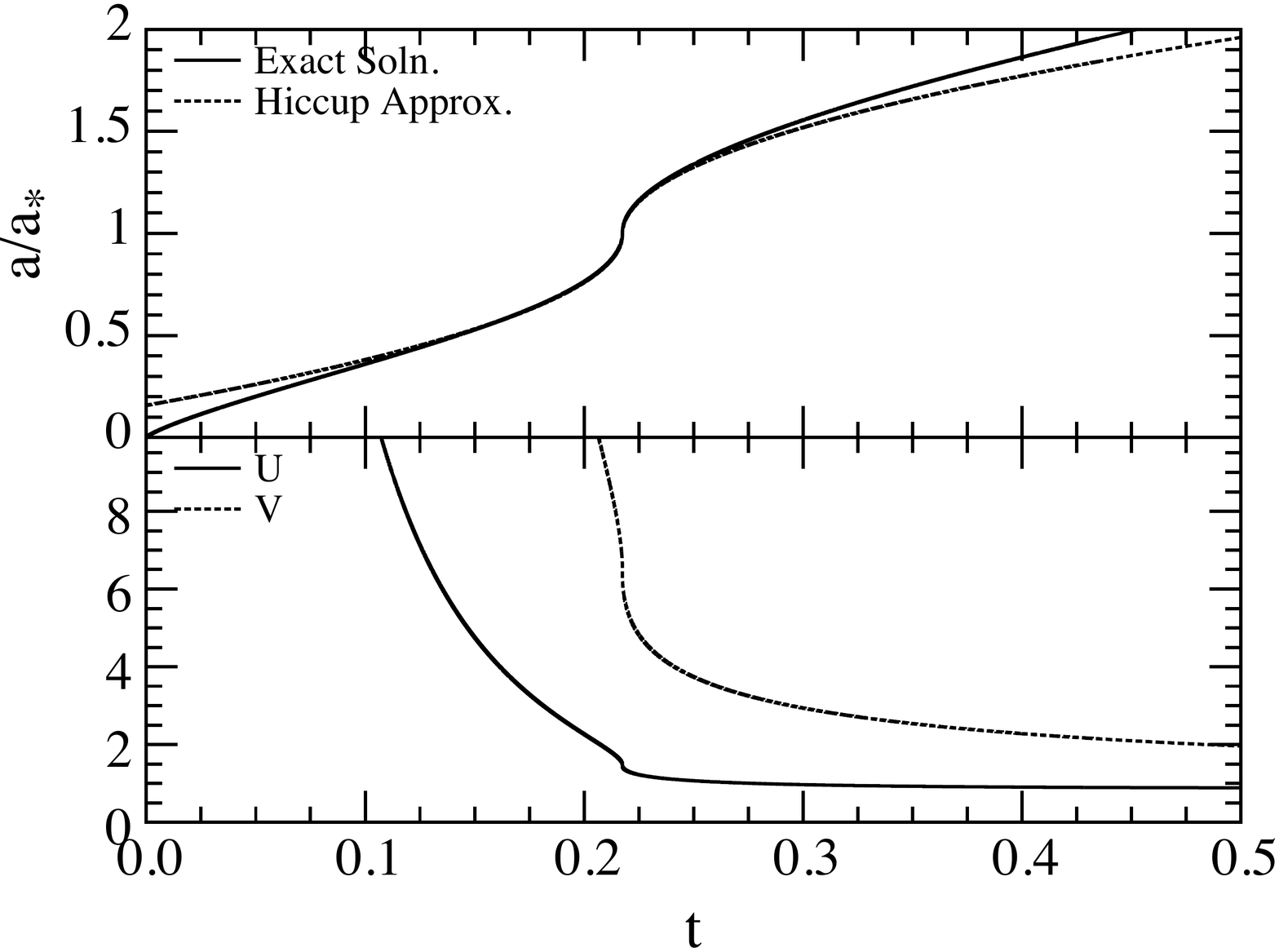}
   \caption{Left: Expansion rate of the universe against density (normalised by the critical density $\rho_*$), for $\kappa = 1$, $w = -1/6$, $k = 0$. \\
   Right Top: Scale factor (normalised by the critical scale factor $a_*$) against time (in reduced Planck units), for the full numerical solution to the Friedman equation, and an approximation close to the critical density. \\
   Right Bottom: Behaviour of the parameters describing the auxiliary metric (for the same universe).}
   \label{Hrho-aUV_2}
\end{figure*}

As shown in table \ref{discontinuities}, in such a universe the expansion rate diverges at a finite density given by the positive root of the denominator of eqn. (\ref{Friedman w/o phi})
\begin{equation}
\bar{\rho}_* = \frac{(1-w)(1-3w)+(1+w)\sqrt{1-42w + 9w^2}}{4|w| (1+3w)}.
\end{equation}
Expanding around this point, to leading order we find
\begin{equation}
H^2 = \frac{8}{3 \kappa} f_w^2 \left( \frac{\delta a}{a_*} \right)^{-2}
\end{equation}
with
\begin{IEEEeqnarray}{rcl}
f_w^2 &=& \Big[ (1+3w)\bar{\rho}_* - 2 + 2 \sqrt{(1+\bar{\rho}_*)(1-w\bar{\rho}_*)^3} \nonumber \\
&&- \frac{6 \bar{k}}{a_*^2}(1-w\bar{\rho}_*) \Big] \frac{1}{(3\bar{\rho}_*(1+w))^2} \nonumber \\
&& \times \frac{(1+\bar{\rho}_*)(1-w\bar{\rho}_*)^2}{((1-w)(1-3w) + 4w(1+3w) \bar{\rho}_*)^2}.
\end{IEEEeqnarray}
This can be solved as in section \ref{loitering phase} to get
\begin{equation}
\frac{a(t)}{a_*} = 1 \pm \sqrt{2 f_w \sqrt{\frac{8}{3 \kappa}} |t - t_*|}. \label{a(t) hiccup}
\end{equation}

Thus at this critical point the universe in some sense undergoes a second big bang, with expansion similar to that of GR with domination by radiation ($a \propto \sqrt{t}$). Figure \ref{Hrho-aUV_2} demonstrates this universe.

Expanding eqn. (\ref{Friedman w/o phi}) for high densities we find
\begin{equation}
H^2 = \frac{4}{3} \frac{| w |^{\frac{3}{2}}}{(1+3w)^2} \rho. \label{hiccup high density Friedman}
\end{equation}
The expansion in the early history of this universe is also GR-like, albeit with a modified density $\tilde{\rho} = \frac{4 | w |^{\frac{3}{2}}}{(1+3w)^2} \rho$, and the spatial curvature can be neglected. The scale factor as a function of time is then
\begin{equation}
\frac{a(t)}{a_*} = \left(\sqrt{\frac{\tilde{\rho}_*}{3}} \frac{3(1+w)}{2} t \right)^{\frac{2}{3(1+w)}} \label{hiccup high density a(t)}
\end{equation}

In order to better understand the dynamics of the universe during this second big bang/`cosmic hiccup' we should consider the horizon between two times $t_- < t_* < t_+$ close to $t_*$:
\begin{equation}
r(t_-,t_+) = \frac{t_+ - t_-}{a_*} + \mathcal{O}((t_\pm-t_*)^2).
\end{equation}
Its well-behaved nature should not be surprising, however, since the nature of the expansion during the `hiccup' has already been noted.

\subsubsection{Length of first life} \label{hiccup time}

An interesting quantity to consider is $t_*$, the time between the birth of the universe and the `hiccup'. Naively, since $\bar{\rho}_* \to \infty$ for $w \to ~ 0,\ -\frac{1}{3}$, it might be expected that $t_* \to 0$ in these limits.

We have
\begin{equation}
t_* = \int_0^{t_*} \mathrm{d}t = \int_0^{a_*} \frac{\mathrm{d} a}{a H(a)} \propto \int_{\rho_*}^\infty \frac{\mathrm{d} \rho}{\rho H(\rho)}. \label{t_* integral}
\end{equation}
Whilst this cannot be evaluated analytically in the general case, certain approximations can be made. For $w \to -\frac{1}{3}$, $\rho_* \propto \frac{1}{1+3w}$, and $H$ is given by eqn. (\ref{hiccup high density Friedman}) and so the behaviour of eqn. (\ref{t_* integral}) can be approximated by $t_* \propto \int_\infty^{\rho_*} \frac{(1+3w)\mathrm{d} \rho}{\rho^{3/2}} = \frac{2(1+3w)}{\rho_*^{1/2}} \propto (1+3w)^{3/2} \to 0$, as expected.

However for $w \to 0$, $\rho_* \to \frac{1}{2|w|}$ and so eqn. (\ref{t_* integral}) behaves: $t_* \propto \int_\infty^{\rho_*} \frac{\mathrm{d} \rho}{|w|^{3/4} \rho^{3/2}} = \frac{2}{|w|^{3/4} \rho_*^{1/2}} \sim |w|^{-1/4} \to \infty$, which defies the naive expectations. This occurs because the expansion rate of the universe heads towards zero at high densities more quickly than $\bar{\rho}_* \to \infty$, as indeed it should if the high density behaviour observed in section \ref{k > 0, w = 0} is to be recovered when $w = 0$.

\subsubsection{Behaviour of the auxiliary metric}

In the high density limit the overall size of the space described by $q$ diverges, but the combined effect of $U$ and $V$ is that $a_q$ remains proportional to the regular scale factor.

At the `hiccup' $U$ and $V$ as functions of $\bar{\rho}$ or $a$ are smooth and well behaved, but as functions of $t$ they of course exhibit a `kink' just as $a(t)$ does (see figure \ref{Hrho-aUV_2}).

Explicit functional forms in these two limits can be found in appendix \ref{k > 0, -1/3 < w < 0, auxiliary metric}.

\begin{figure*}[htbp]
   \includegraphics[width=\columnwidth]{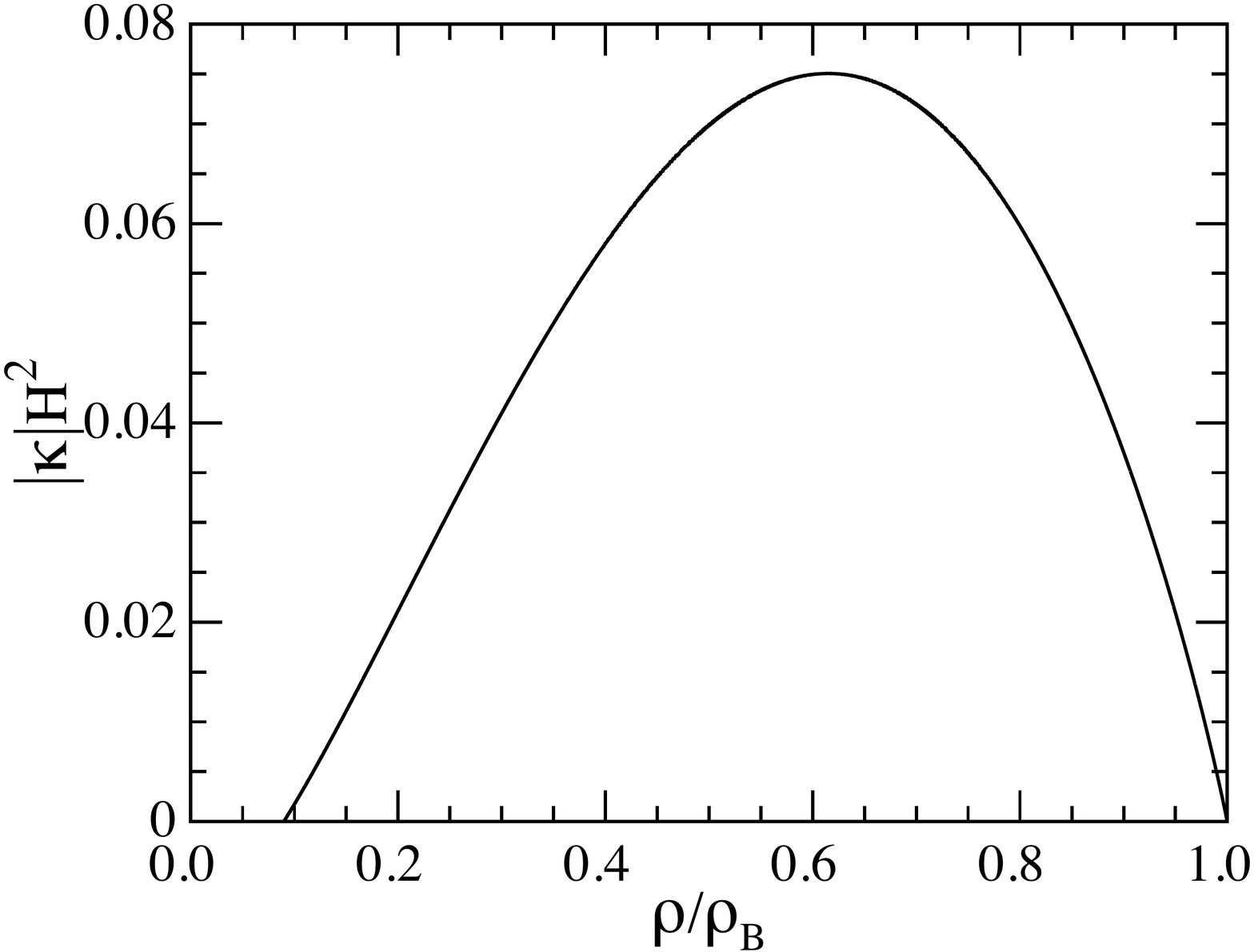}
   \includegraphics[width=\columnwidth]{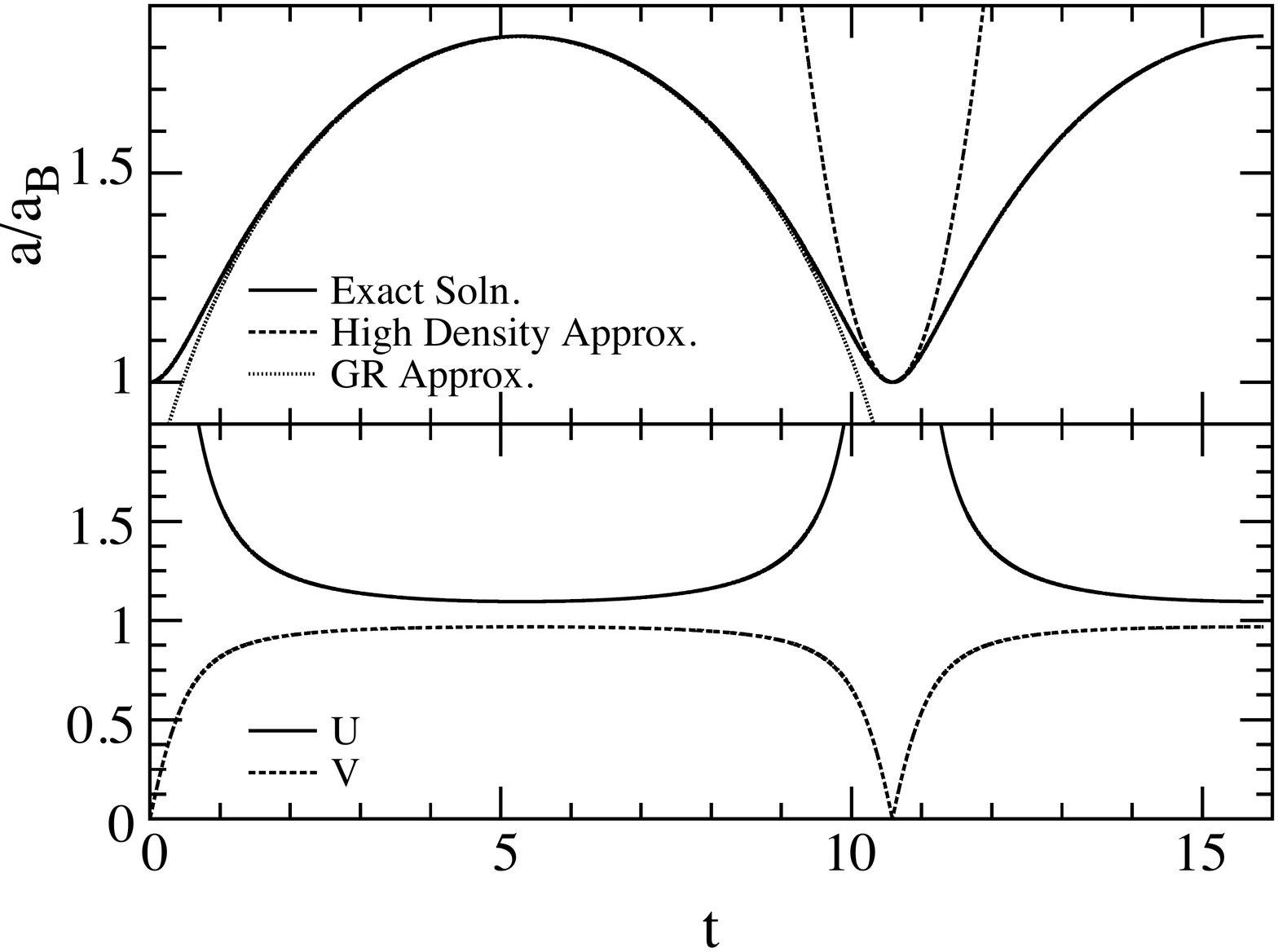}
   \caption{Left: Expansion rate of the universe against density (normalised by the maximum density $\rho_B = - \kappa^{-1}$), for $\kappa = -1$, $w = 1/3$, $k/a_B^2 = 1/10$. \\
   Right Top: Scale factor (normalised by the minimum scale factor $a_B$) against time (in reduced Planck units), for the full numerical solution to the Friedman equation, and approximations in two phases: close to the maximum density, and GR-like low-density limit. \\
   Right Bottom: Behaviour of the parameters describing the auxiliary metric (for the same universe).}
   \label{Hrho-aUV3}
\end{figure*}

\subsection{Oscillating universes} \label{oscillating universes}

In \cite{Banados and Ferreira, Avelino and Ferreira} and \cite{Cho et al} it is noted that for $\kappa < 0$, $w > 0$ the universe undergoes a `regular' bounce (as described below) in which, in a finite time, a collapsing universe reaches a maximum density and then expands. In GR, universes with positive spatial curvature can undergo a bounce in the opposite direction.

The addition of spatial curvature to Eddington cosmology thus leads to the interesting prospect of oscillatory behaviour. Figure \ref{Hrho-aUV3} demonstrates such an oscillatory universe.

\subsubsection{Amplitude of the oscillations} \label{oscillation amplitude}

The maximum density, $\rho_B = - \kappa^{-1}$, is independent of $w$ and $k$, and is caused by the $(1+\bar{\rho})$ factor in the numerator of eqn. (\ref{Friedman w/o phi}). The minimum density, on the other hand, does depend on $w$ and $k$ (and $\kappa$), and is the solution to $G = 0$, which explicitly is
\begin{align}
&(1+3w) \bar{\rho} - 2 + 2\sqrt{(1+\bar{\rho})(1-w\bar{\rho})^3} \nonumber \\
&- 6\bar{k}a^{-2}(1-w\bar{\rho}) = 0.
\end{align}
Unfortunately this cannot be solved analytically in the general case, however if we assume that the universe has expanded sufficiently before expansion halts then we can solve this in the GR regime to get
\begin{equation}
\frac{a_{max}}{a_B} = \left(\frac{3k}{a_B^2}\right)^{-\frac{1}{1+3w}}, \  \frac{\rho_{min}}{\rho_B} = \left(\frac{3k}{a_B^2}\right)^{\frac{3(1+w)}{1+3w}} \label{GR oscillation amplitude}
\end{equation}

\begin{figure}[htbp]
   \includegraphics[width=\columnwidth]{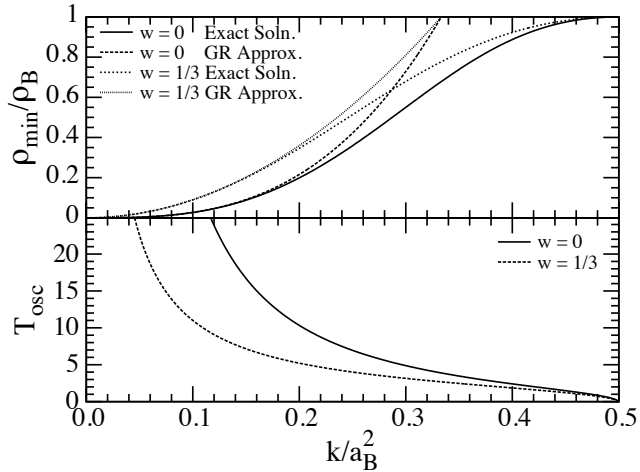}
   \caption{Top: Amplitude (minimum density, normalised by maximum density) of the oscillations described in section \ref{oscillating universes} against spatial curvature of the universe, for $\kappa = -1$ and various values for the equation of state parameter. Full numerical solutions, as well as approximations in the GR regime from eqn. (\ref{GR oscillation amplitude}).\\
   Bottom: Period of the oscillations (in reduced Planck units) for the same universes.}
   \label{rmax_tosc_3}
\end{figure}

From figure \ref{rmax_tosc_3} we see that this is indeed a good approximation for $k/a_B^2 \lesssim 0.2$.

Close to the maximum density we have
\begin{equation}
H^2 = \frac{8}{3 |\kappa|} \left(1 + \frac{2 \bar{k}}{a_B^2} \right) \frac{\delta a}{a_B}, \label{oscillating Friedman}
\end{equation}
which is solved to give
\begin{equation}
\frac{a(t)}{a_B} = 1 + \frac{2}{3 |\kappa|} \left(1 + \frac{2 \bar{k}}{a_B^2} \right) |t - t_B|^2. \label{a(t) bounce}
\end{equation}
Note that this is independent of the type of matter which fills the universe (just as in section \ref{loitering phase}), and so we understand the expansion is being driven by $\kappa$.

Bounds can be placed on $k$ by examining the Friedman equation in these two limits (since the behaviour of eqn. (\ref{Friedman w/o phi}) as a function of $\bar{\rho}$ and $k$ is smooth) and requiring $H^2 \geq 0$. The GR-like regime gives a lower bound on $k$ whilst eqn. (\ref{oscillating Friedman}) gives an upper bound, which combine to give
\begin{equation}
0 < \frac{k}{a_B^2} < \frac{1}{2 |\kappa|}.
\end{equation}

\subsubsection{Period of the oscillations}

In principle the period of the oscillations can be calculated from
\begin{equation}
T_{osc} = 2\int_0^{T_{osc}/2} \mathrm{d} t = 2 \int_{a_B}^{a_{max}} \frac{\mathrm{d} a}{a H(a)}. \label{oscillation period}
\end{equation}
In practice this integral cannot be performed analytically (or good enough analytical approximations made) and so must be evaluated numerically, as shown in figure \ref{rmax_tosc_3}.

We see that $T_{osc}(w = \frac{1}{3}) < T_{osc}(w = 0)$, which makes sense since the amplitude of the oscillations for the former case is always smaller than in the latter. Also note that $T_{osc} \to 0$ for $k/a_B^2 \to 1/2$. From eqn. (\ref{oscillating Friedman}) we see that the expansion rate goes to zero at this point, and so from section \ref{hiccup time} we might expect that $T_{osc} \to \infty$, however in this case the extra size by which the universe must expand also heads to zero (and clearly does so more quickly than the expansion rate).

\subsubsection{Behaviour of the auxiliary metric}

Close to the maximum density the spatial size decreases to zero as the universe reaches is minimum size, whilst the temporal component of the auxiliary metric diverges. See appendix \ref{k < 0, w > 0, auxiliary metric} for approximate functional forms for these.

\section{Dynamics of a universe with a Scalar Field} \label{w phi}

We now consider the cosmology when a scalar field is introduced; the ordinary matter content, the cosmological constant, and the spatial curvature are considered negligible. Written in terms of $\phi \text{ and } V(\phi)$ the FRW equation is
\begin{widetext}
\begin{align}
H = \Bigg[&\pm \left(1- \frac{\kappa}{2}\dot{\phi}^2 + \kappa V \right) \sqrt{\left(1 + \frac{\kappa}{2} \dot{\phi}^2 + \kappa V \right) \left( \frac{1}{3} \left\{\dot{\phi}^2 - V - \frac{1}{\kappa} \left[1 + \sqrt{\left(1 + \frac{\kappa}{2} \dot{\phi}^2 + \kappa V\right) \left(1- \frac{\kappa}{2}\dot{\phi}^2 + \kappa V \right)^3} \right] \right\} \right)} \nonumber \\ &  - \frac{\kappa}{2} \left(1 + \frac{\kappa}{2} \dot{\phi}^2 + \kappa V \right) V' \dot{\phi} \Bigg] \frac{1}{1 + 2 \kappa V + \kappa^2 \left( \frac{1}{2} \dot{\phi}^4 + V^2 \right)}. \label{Friedman w phi}
\end{align}
\end{widetext}

The positive sign is taken, as this gives a universe which is expanding in the GR regime. The conservation equation governing the scalar field is
\begin{equation}
\ddot{\phi} + 3 H \dot{\phi} + V'(\phi) = 0.
\end{equation}
We restrict ourselves to an exponential potential for the scalar field
\begin{equation}
V(\phi) = V_0 \,\mathrm{exp}(-\lambda \phi)
\end{equation}
with $V_0,\,\lambda > 0$. Unless explicitly included, the parameter describing deviations from GR ($\kappa$) is set to one.

\subsection{Loitering behaviour} \label{loitering}

\begin{figure*}[tbp]
   \centering
   \includegraphics[width=\columnwidth]{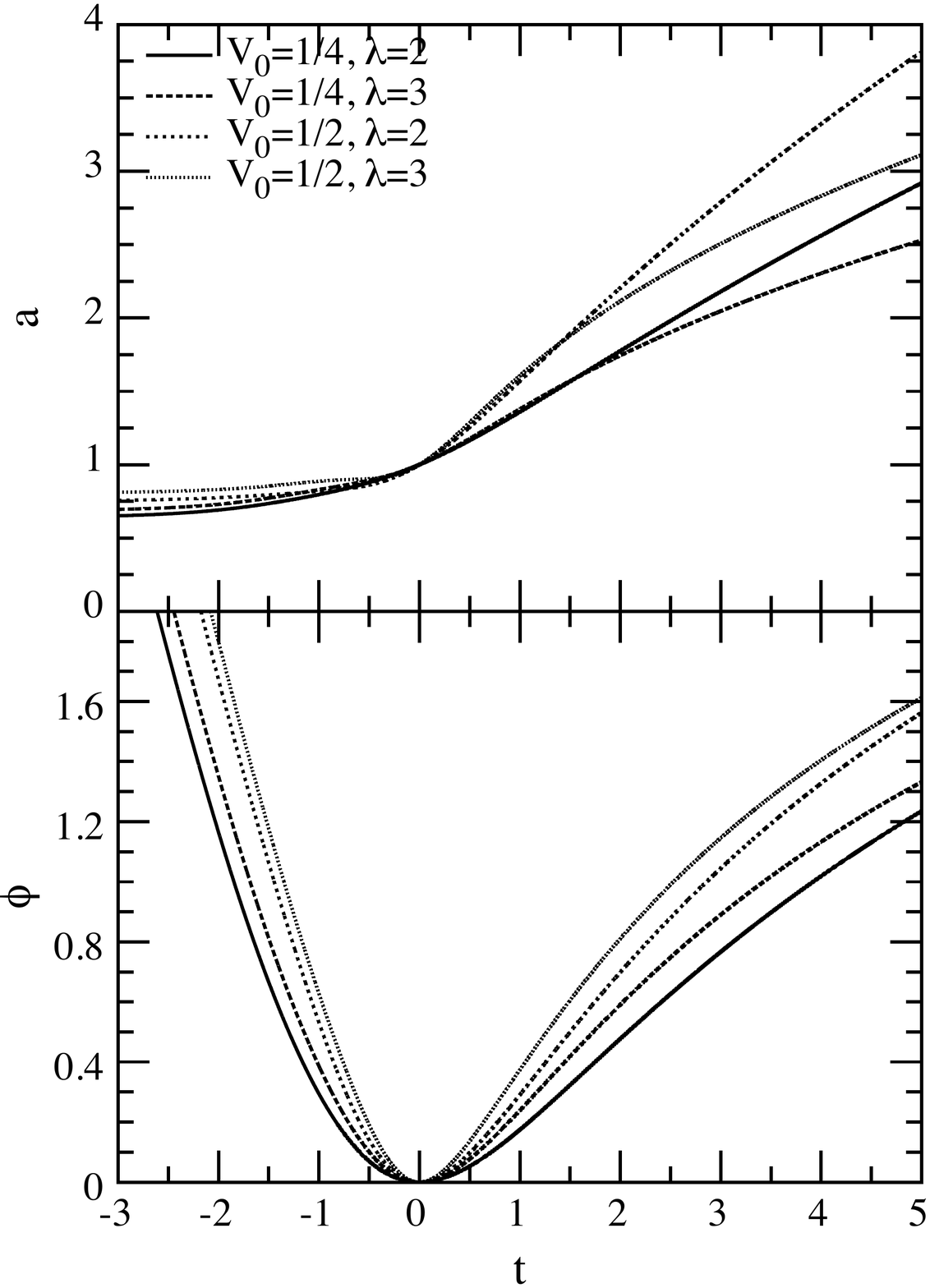}
   \includegraphics[width=\columnwidth]{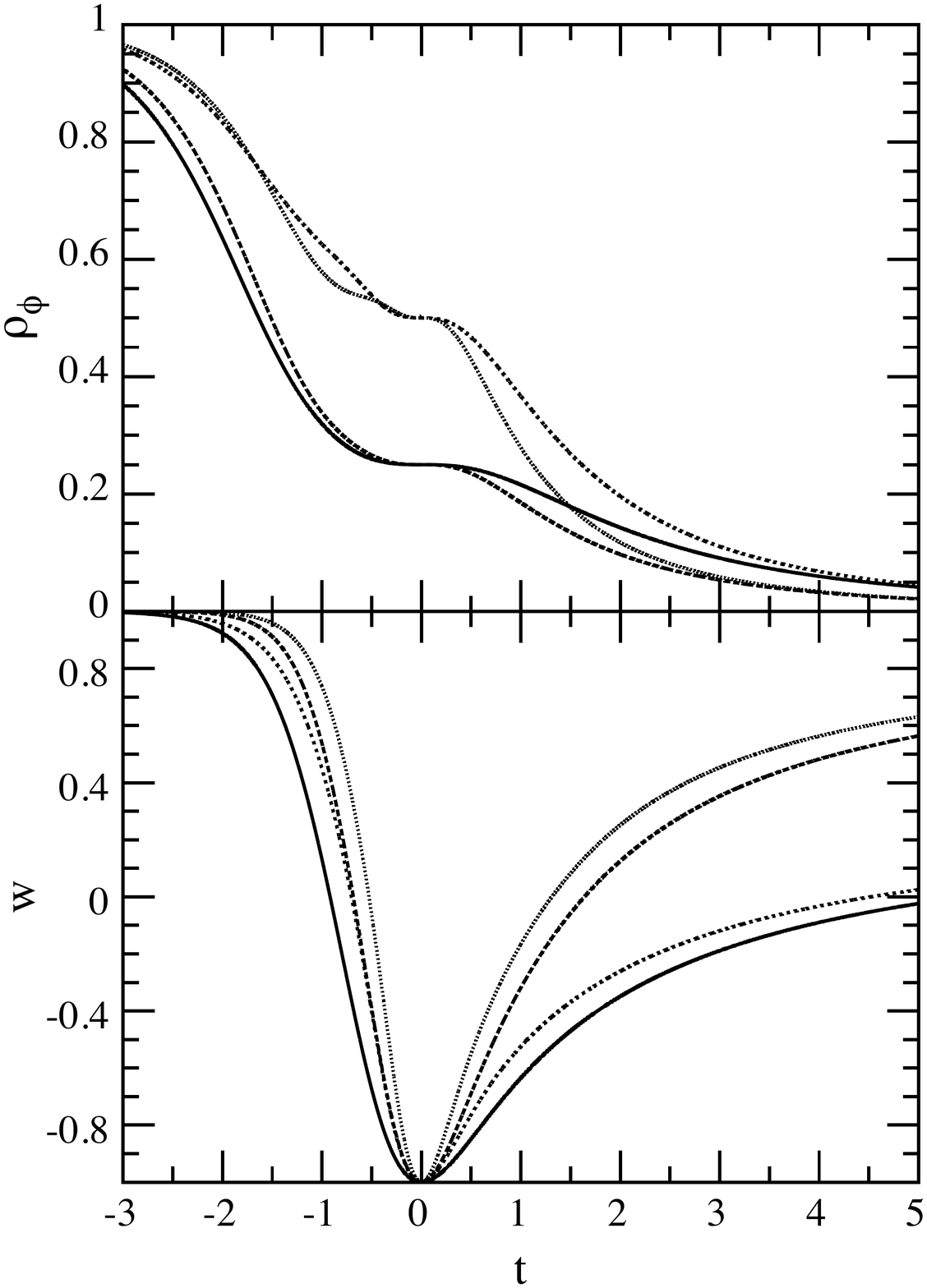}
   \caption{Scale factor (top left), scalar field (bottom left), energy density (top right) and equation of state parameter ($w \equiv P/\rho$) (bottom right) as a function of time (in reduced Planck units) for various values of the free parameters.}
   \label{a_phi_rho_w_ordinary}
\end{figure*}

For $\kappa > 0$ numerical solutions of eqn. (\ref{Friedman w phi}) show qualitatively similar behaviour to the case without scalar modes, viz. `loitering' of the scale factor around a minimum value followed by a transition to GR-like power-law expansion. See figure \ref{a_phi_rho_w_ordinary}, top left panel.

The solution for $\phi(t)$ also shows two regimes with a transition between them (figure \ref{a_phi_rho_w_ordinary}, bottom right panel). The GR regime shows the expected logarithmic time dependence \cite{Ferreira+Joyce}, whilst the Eddington/loitering regime shows $\phi \propto -t$.

\subsubsection{Initial conditions and free parameters}

It is appropriate first to briefly discuss the initial conditions used  to produce the solutions plotted above as well as the effect on these that is had by altering the free parameters of the system.

As can be seen from figure \ref{a_phi_rho_w_ordinary}, the chosen conditions are
\begin{equation}
a(t=0) = 1; \quad \phi(0) = 0; \quad \dot{\phi}(0) = 0.
\end{equation}
The first is uncontroversial since the scale factor can always be rescaled by a constant. The absolute value of the scalar field only appears in the form for its potential energy, and thus adjusting the field by a constant, which is the action of the second i.c., only has the effect of changing $V_0$. The theory is obviously invariant under a constant time shift and the third i.c. just fixes this. The reader may be worried that using $\dot{\phi}=0$ as a i.c. eliminates from consideration any solutions which show monotonically increasing field values, however such solutions would not exhibit the maximum density and loitering behaviour (since if $\phi$ is unbounded from below the density will also be unbounded, and if $\phi$ tends to some constant value, the pressure would be negative and hence not satisfy $P = \kappa^{-1}$).

All that's left is to describe the potential, which leaves two free parameters: $V_0$ and $\lambda$. From the figures thus far presented we can roughly evaluate the effect of making the potential steeper (increasing $\lambda$) or deeper (increasing $V_0$).

The late time (GR-like) behaviour is primarily governed by the value of $\lambda$, whereas the early time behaviour is influenced by $V_0$. The reason for the latter is that the energy density of the field is already fixed as $t \to -\infty$, and picking a value for $V_0$ fixes it at $t = 0$ (or wherever the transitionary period is located), and thus the behaviour in between these two points will be greatly influenced by the size of the change in the energy density.

Furthermore one can roughly say that steeper and deeper potentials lead to quicker transitions from the loitering to the GR phase.

\subsubsection{Loitering regime}

Let us investigate this loitering behaviour in more detail. As in the case with ordinary matter the loitering regime corresponds to $P \to \kappa^{-1}$ (as $t \to - \infty$); we also see that $\phi \to \infty$ and hence $V(\phi) \to 0$. Thus one has $\rho \approx P$ and $\phi \to -\sqrt{\frac{2}{\kappa}}t + \text{const}$.

Evaluating the Friedman equation in this regime, one finds
\begin{equation}
H(P = \kappa^{-1}) = \frac{2 (1 + \kappa \rho) V'}{3 \kappa \dot{\phi}^3} = \frac{1}{3}\sqrt{2 \kappa} V'.
\end{equation}
Therefore the equation of motion for the scalar field reduces to $\ddot{\phi} + 3 V' = 0$. The solution (setting $\kappa = 1$) is
\begin{align}
\phi(t) &= -\sqrt{2} (t - t_0) + \frac{1}{\lambda} \left[ 2\ln ( 12 V_0 + \mathrm{e}^{\sqrt{2} \lambda (t-t_0)} ) -4 \ln 2 \right]\\
a(t) &\approx a_{min} \exp\left(\frac{16 V_0}{3} \exp \sqrt{2} \lambda (t-t_0) \right).
\end{align}

The dynamic nature of the scalar field adjusts the loitering behaviour of the scale factor from the simple exponential behaviour seen in the case of regular matter to the more extreme double exponential.

\subsubsection{Transitionary Regime}

Whereas in the case of ordinary matter fields the evolution is well described without having to denote a separate transitionary regime (between the high-density loitering behaviour and the low-density GR-like regime), that is not the case here.

The transition from linearly decreasing field values to logarithmically increase is naturally centred on $\dot{\phi} = 0$. Hence during the transition $P \approx - \rho$: the scalar field has the equation of state of a cosmological constant. This is readily observed in plots of $\rho(t)$ which show that the density is stationary around the transition. See figure \ref{a_phi_rho_w_ordinary}, top right panel.

The bottom right panel of this figure also confirm this, as well as the above comment that $P \approx \rho$ in the loitering regime ($w \to 1$ as $t \to -\infty$, regardless of the values of the free parameters).


\subsubsection{Density Scaling}

\begin{figure}[tbp]
   \centering
   \includegraphics[width=\columnwidth]{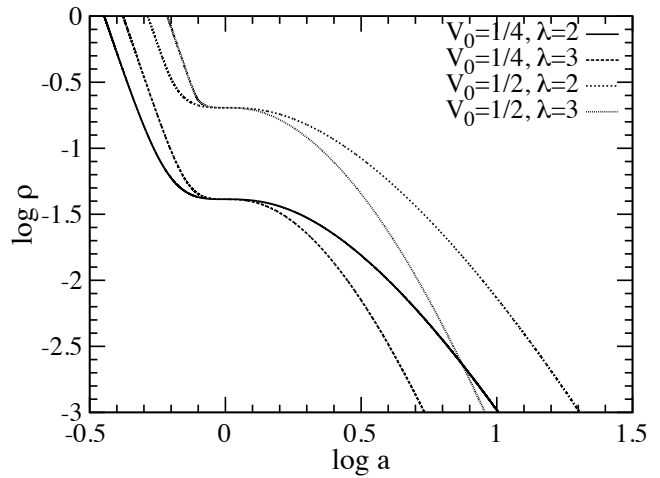}
   \caption{Energy density as a function of scale factor for various values of the free parameters.}
   \label{log(rho-a)_ordinary}
\end{figure}

Finally it is interesting to look at the plot of $\ln \rho$ against $\ln a$ (figure \ref{log(rho-a)_ordinary}). This shows that during the loitering phase the density scales as $\rho \propto a^{-6}$, which agrees with the fact that $w \approx 1$ - the scalar field is, to use the terminology used in \cite{Ferreira+Joyce}, undergoing \emph{kination} (kinetic energy domination). This is followed by the expected cosmological constant-like transitionary phase in which the density is stationary. The field energy density shows the expected scaling in the GR phase ($\rho \propto a^{-\lambda^2} \text{ if } \lambda < \sqrt{6} \text{ and } \rho \propto a^{-6} \text{ otherwise}$).

Figure 1 from \cite{Ferreira+Joyce} shows the evolution of the energy density of a scalar field in a GR-universe populated by the aforementioned scalar field (with $\lambda = 4$) as well as radiation and cold  matter. The scalar field is initially dominant and scales accordingly ($\rho \propto a^{-6}$), becomes sub-dominant and scales like a constant, before finally tracking (i.e. scaling as) the cold matter ($\rho \propto a^{-3}$) which is then dominant.

This bears a remarkable resemblance to figure \ref{log(rho-a)_ordinary}, except for two major differences:
\begin{enumerate}
\item the behaviour in the Eddington case is reversed. The `natural' scaling (i.e. that which one would expect in GR based on the value of $\lambda$) is preceded by a period in which the scaling is fixed (by something other than $\lambda$); \\
\item in the Eddington case such behaviour is achieved without another fluid for the field to track.
\end{enumerate}

\subsection{Expand-contract-expand behaviour} \label{expconexp}

As noted in section \ref{Friedman eqn. general}, the two solutions for the Hubble parameter are not in general simply of the same magnitude but with opposite sign. This leads to the possibility that there are universes which at some point are stationary, and then instead of contracting and retracing their evolutionary history in reverse, they contract but trace a new history.\footnote{It should be noted that if one allows $\dot{\phi}$ (and hence $\rho_\phi$) to be discontinuous, then the standard GR behaviour of symmetry around a bounce is permitted as well.}

In fact since the solutions being investigated are expanding both in their early and late histories, they would undergo expansion, followed by contraction, followed by expansion.

\begin{figure}[tbp]
   \centering
   \includegraphics[width=\columnwidth]{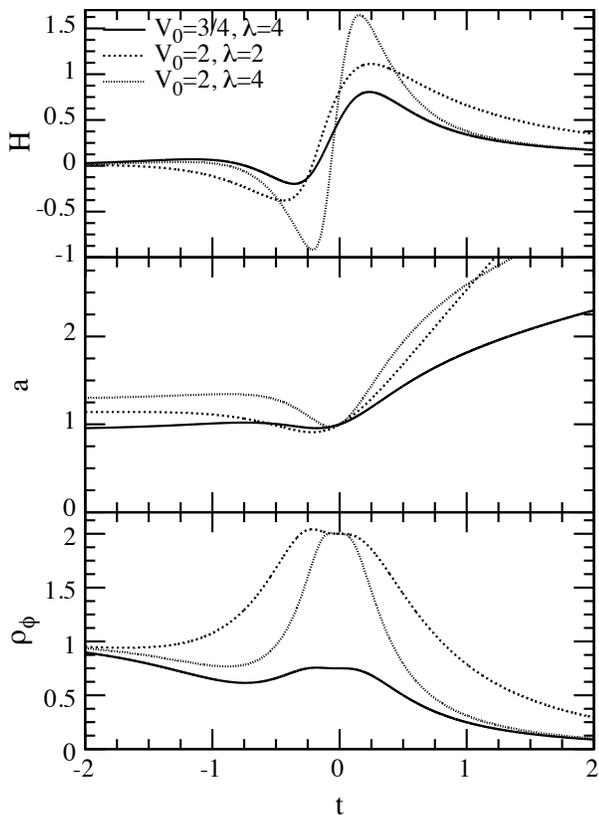}
   \caption{Hubble expansion parameter (top), scale factor (middle), and energy density (bottom) as a function of time (in reduced Planck units) for various values of the free parameters, showing universes which expand-contract-expand.}
   \label{fig_expconexp}
\end{figure}

From figure \ref{fig_expconexp} we see that such behaviour can be achieved both by making the potential steeper and by making it deeper. The latter can easily be understood since $V_0$, which controls the depth of the potential, also controls the energy density of the scalar field at its transitionary point. Thus choosing a sufficiently large $V_0$ will ensure that the density at this point is larger than the density in the distant past (loitering phase), and hence the universe must undergo contraction (see figure \ref{fig_expconexp}, bottom panel).

That steeper potentials induce this behaviour can be understood mathematically from the fact that the `$B$' term in the Friedman equation (eqns. \ref{Friedman}, \ref{B}) is proportional to $V'$, and hence to $\lambda$. Physically, perhaps one can imagine the cause to be that a potential which varies quickly enough with $\phi$ leads the scalar field to `overshoot' the value which would ensure energy density monotonically decreases with time; this larger density would then be associated with a contraction.

It is interesting to note that plots of the field value, or of the equation of state parameter do not obviously show any effect of the period of contraction or differ qualitatively from the corresponding plots for free parameter values which show the ordinary behaviour (figure \ref{a_phi_rho_w_ordinary}).

Unfortunately we have yet to determine the critical values of $V_0$ and $\lambda$ which guard the transition into such behaviour.

\section{Conclusion} \label{conclusion}

In this paper we have shown that the singularity-avoiding `loitering' behaviour observed for $\kappa > 0$ and domination by a perfect fluid is upheld for $w > 0$ and that this result is not affected by the presence of spatial curvature. This behaviour shows remarkable similarities to the behaviour in GR with a cosmological constant $\Lambda = 8 \kappa^{-1}$. Similar behaviour is also seen in the case of domination by a scalar field with an exponential potential. The scaling of the density with the scale factor in such universes is reminiscent of that observed in a GR universe with such a scalar field and (initially more slowly scaling) ordinary matter.The similarities to GR in each of these cases deserve further investigation which will shed light on the high density behaviour of Eddington gravity.

The (past) singularity avoiding behaviour of this theory should be compared to that examined in \cite{Capozziello} in which higher derivative terms in $f(R)$ theories lead to future finite time singularity avoidance. The non-linear structure of the Eddington gravity action introduces such terms and so we can perhaps qualitatively understand that this leads to singularity avoidance in a similar manner.

Finally as was noted in section \ref{Friedman eqn. general}, in the low density limit ($\rho \ll \kappa^{-1}$, with $\kappa^{-1} \gtrsim 10^{18}$ kg m$^{-3}$ \cite{Avelino Constraints}) GR is recovered and so the theory is indistinguishable from $\Lambda$CDM (or similar dark energy models) in this cosmological arena.

For both perfect fluids and scalar fields we have also demonstrated novel behaviour unlike that seen in GR. In the former case we have the `cosmic hiccup' which is observed for $\kappa > 0$, $-\frac{1}{3} < w < 0$, as well as the oscillatory universes for $\kappa < 0$, $w > 0$, and positive spatial curvature. In the latter case we have shown that especially deep or steep scalar field potentials lead the size of the universe oscillating whilst transitioning between the loitering and GR-like phases.

There appears no initial reason why the singularity avoiding behaviour observed for $\kappa < 0$ cannot also exist for a scalar field dominated universe. As explained above, symmetry around the bounce would be permitted in this case ($\dot{\phi} = 0$ at the bounce).Furthermore the interplay between the scalar field and regular matter fields could lead to interesting behaviour of the scalar field, as is the case in GR \cite{Ferreira+Joyce} and remains to be studied in detail.

\begin{acknowledgments}
This research is supported by STFC, Oxford Martin School and BIPAC.  MB was partially supported by Fondecyt (Chile) Grants \#1100282 and \# 1090753.
\end{acknowledgments}

\appendix

\section{Regions in which discontinuities appear} \label{appendix discontinuities}

\begin{figure*}[htbp]
   \centering
   \includegraphics[width=\textwidth]{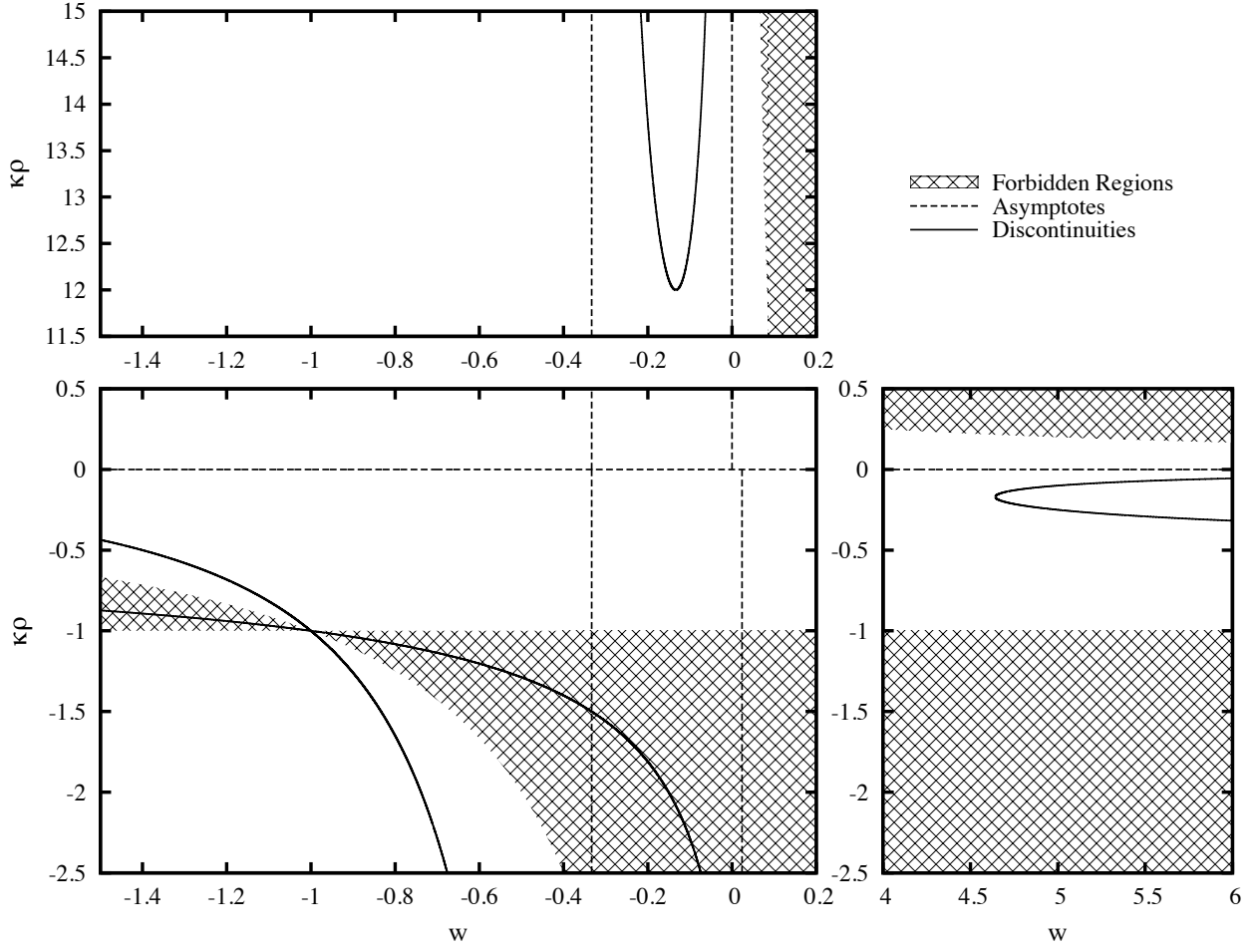}
   \caption{Hatched regions are those forbidden due to the square root in $G$; solid lines indicate the solutions to $F = 0$ and so represent the locations of discontinuities in $H^2$; dashed lines indicate the asymptotes to which the solid lines tend.}
   \label{figure discontinuities}
\end{figure*}

Figure \ref{figure discontinuities} shows, for a universe dominated by a single perfect fluid, the locations in $(\bar{\rho},w)$ parameter space of discontinuities in $H^2$. There are no discontinuities at all for $0.02 \approx \frac{1}{3}(7-4\sqrt{3}) < w < \frac{1}{3}(7+4\sqrt{3}) \approx 4.64$, and outside of this range they are given by
\begin{equation}
\bar{\rho}_* = -\frac{(1-w)(1-3w)\pm(1+w)\sqrt{1-42w + 9w^2}}{4w (1+3w)}.
\end{equation}

Also shown are the regions which are forbidden, due to the square root in $G$ (and the requirement that $H^2$ is real). Combining these data we arrive at table \ref{discontinuities}.

\section{Approximate functional forms for the behaviour of the auxiliary metric}

\subsection{The case $\kappa > 0$, $w = 0$} \label{k > 0, w = 0, auxiliary metric}

Defining $\bar{\rho}(a = a_0) = \bar{\rho}_0$, during the loitering phase we find
\begin{align}
U &= \frac{1}{\sqrt{\bar{\rho}_0}} \left( \frac{\delta a}{a_0} \right)^{\frac{3}{2}} = \frac{1}{\sqrt{\bar{\rho}_0}} \exp\left( \frac{3}{2} \sqrt{\frac{8}{3\kappa}} (t-t_0) \right) \\
V &= \sqrt{\bar{\rho}_0} \left( \frac{\delta a}{a_0} \right)^{-\frac{3}{2}} = \sqrt{\bar{\rho}_0} \exp\left( - \frac{3}{2} \sqrt{\frac{8}{3\kappa}} (t-t_0) \right) \\
a_q &= a_0 \sqrt{\bar{\rho}_0} \left( \frac{\delta a}{a_0} \right)^{-\frac{3}{2}} = a_0 \sqrt{\bar{\rho}_0} \exp \left( -\frac{3}{2} \sqrt{\frac{8}{3\kappa}} (t-t_0) \right).
\end{align}

\subsection{The case $\kappa > 0$, $-\frac{1}{3} < w < 0$} \label{k > 0, -1/3 < w < 0, auxiliary metric}

In the high density limit, for the early history of this universe, we have
\begin{equation}
U = |w|^{\frac{3}{2}} \bar{\rho}, \quad
V = |w|^{\frac{1}{2}} \bar{\rho}, \quad
a_q = a |w|^{-\frac{1}{2}}.
\end{equation}
Where $a$ is given by (\ref{hiccup high density a(t)}) and thus $\bar{\rho} = ~ \bar{\rho}_* \left(\sqrt{\frac{\tilde{\rho}_*}{3}} \frac{3(1+w)}{2} t \right)^{-2}$.

The functional forms close to the critical point are
\begin{align}
U &= U_*\bigg[ 1 -\frac{3 \bar{\rho}_* (1+w)}{2} \left( \frac{1}{1+ \bar{\rho}_*} + \frac{3w}{1-w\bar{\rho}_*} \right)\frac{\delta a}{a_*} \bigg] \\
V &= V_*\bigg[ 1 -\frac{3 \bar{\rho}_* (1+w)}{2} \left( \frac{1}{1+ \bar{\rho}_*} - \frac{w}{1-w\bar{\rho}_*}  \right)\frac{\delta a}{a_*} \bigg] \\
a_q &= {a_q}_*\bigg[ 1+ \bigg[1 -\frac{3 \bar{\rho}_* (1+w)}{2} \left( \frac{1}{1+ \bar{\rho}_*} + \frac{w}{1-w\bar{\rho}_*}  \right)\bigg]\frac{\delta a}{a_*} \bigg]
\end{align}
and $\frac{\delta a}{a_*}$ can be read off from (\ref{a(t) hiccup}).

\subsection{The case $\kappa < 0$, $w > 0$} \label{k < 0, w > 0, auxiliary metric}

Close to the maximum density we have
\begin{align}
U &= (1+w) \sqrt{\frac{a_B}{3 \delta a}} = (1+w) \sqrt{\frac{2}{| \kappa |} \left( 1 + \frac{2 \bar{k}}{a_B^2} \right) }^{-1} |t - t_B|^{-1}\\
V &= (1+w) \sqrt{\frac{3 \delta a}{a_B}} = (1+w) \sqrt{\frac{2}{| \kappa |} \left( 1 + \frac{2 \bar{k}}{a_B^2} \right) } |t - t_B|\\
a_q &= 3 \delta a = \frac{2 a_B}{| \kappa |} \left( 1 + \frac{2 \bar{k}}{a_B^2} \right)  |t - t_B|^2.
\end{align}

\end{document}